\pdfoutput=1

\documentclass[sigconf]{acmart}
\AtBeginDocument{%
  \providecommand\BibTeX{{%
    \normalfont B\kern-0.5em{\scshape i\kern-0.25em b}\kern-0.8em\TeX}}}

\setcopyright{None}
\acmDOI{}

\acmConference[]{}{}{}
\acmPrice{}
\acmISBN{}

\begin{document}
\title{Collaborative Cloud Computing Framework for Health Data with Open Source Technologies}

\author{Fatemeh Rouzbeh}
\affiliation{%
  \institution{Purdue University}
  \city{West Lafayette}
  \country{USA}}
  \email{frouzbeh@purdue.edu}

\author{Ananth Grama}
\affiliation{%
  \institution{Purdue University}
  \city{West Lafayette}
  \country{USA}}
  \email{ayg@purdue.edu}

\author{Paul Griffin}
\affiliation{%
  \institution{Purdue University}
  \city{West Lafayette}
  \country{USA}}
  \email{griff200@purdue.edu}

\author{Mohammad Adibuzzaman}
\affiliation{%
  \institution{Purdue University}
  \city{West Lafayette}
  \country{USA}}
  \email{madibuzz@purdue.edu}


\begin{abstract}
  \label{abstract}
The proliferation of sensor technologies and advancements in data collection methods have enabled the accumulation of very large amounts of data. Increasingly, these datasets are considered for scientific research. However, the design of the system architecture to achieve high performance in terms of parallelization, query processing time, aggregation of heterogeneous data types (e.g., time series, images, structured data, among others), and difficulty in reproducing scientific research remain a major challenge. This is specifically true for health sciences research, where the systems must be i) easy to use with the flexibility to manipulate data at the most granular level, ii) agnostic of programming language kernel, iii) scalable, and iv) compliant with the HIPAA privacy law. In this paper, we review the existing literature for such big data systems for scientific research in health sciences and identify the gaps of the current system landscape. We propose a novel architecture for software-hardware-data ecosystem using open source technologies such as Apache Hadoop, Kubernetes and JupyterHub in a distributed environment. We also evaluate the system using a large clinical data set of 69M patients. 
 
\end{abstract}

\begin{CCSXML}
<ccs2012>
   <concept>
       <concept_id>10010520.10010521.10010537.10003100</concept_id>
       <concept_desc>Computer systems organization~Cloud computing</concept_desc>
       <concept_significance>500</concept_significance>
       </concept>
   <concept>
       <concept_id>10010405.10010444.10010447</concept_id>
       <concept_desc>Applied computing~Health care information systems</concept_desc>
       <concept_significance>500</concept_significance>
       </concept>
 </ccs2012>
\end{CCSXML}

\ccsdesc[500]{Computer systems organization~Cloud computing}
\ccsdesc[500]{Applied computing~Health care information systems}

\keywords{Big Data, Healthcare, Cloud Computing, Hadoop, Kubernetes, JupyterHub.}

\iffalse
\begin{teaserfigure}
  \includegraphics[width=\textwidth]{sampleteaser}
  \caption{Seattle Mariners at Spring Training, 2010.}
  \Description{Enjoying the baseball game from the third-base
  seats. Ichiro Suzuki preparing to bat.}
  \label{fig:teaser}
\end{teaserfigure}
\fi
\maketitle
%Temp 
%\thispagestyle{empty}

\section{Introduction}
\label{intro}
%\IEEEPARstart{A}
According to the \emph{International Data Corporation (IDC)}\cite{reinsel2018digitization}, the global datasphere including finance, manufacturing, healthcare, media and entertainment  will grow from 33 zettabytes in 2018 to 175 zettabytes by 2025. The growing data is shifting the core of the enterprise data repository from traditional datacenters to cloud datacenters (both public and private). According to the IDC, 49\% of the world's stored data will reside in public cloud environments by 2025, up from 25\% in 2019. 

Although health data currently makes up roughly 15\% of the datasphere \cite{reinsel2018digitization}, it is predicted that it will increase at the fastest rate, and outmatch the media and entertainment industries by 2025. Consequently, the need for having an integrated cloud based system to store, manage, and analyze health data in a large scale computing environment is critical \cite{corish_2018}.   

One major motivation for developing curated data repositories and analytics frameworks is widespread recognition of the potential of data analytics solutions to solve critical health science research problems. Several solutions have been provided in various application domains including health sciences and computational biology \cite{murphy2010serving,hripcsak2015observational,murphy2017grappling,kohane2011translational,GS2017Rcupcake,Margolis2014BD2K,ohno2011idash,sittig2012survey,scheufele2014transmart,patel2016database, santana2015towards, cohen2017scientific, li2019cumulus, debauche2017cloud}, materials design \cite{curtarolo2013high}, and mitigating natural disasters~\cite{schnase2017merra,hashem2015rise}. These first generation systems typically collect, curate, and present data in a mediated schema, or federate access across databases using conventional web services technologies. Although these systems have become commonplace, there is an increasing need for higher-level functionality support within these systems. 
For example, reproducibility of scientific research has gained much attention from the scientific community due to the increasing rate of retracted scientific research and proliferation of research articles without strong validation strategies including external validity \cite{ioannidis2005most,gehr2006fading,ioannidis2005contradicted,open2015estimating, freire2012computational}. This is specifically true for health sciences where the safety and efficacy trade-offs are studied with meticulous scrutiny due to regulatory reasons and scientific studies generally go through multiple levels of reproducibility, from animal studies to human trials, before a new treatment strategy is recommended for clinical practice. One general way to validate the result is through replication by re-conducting the entire study. This is in general not feasible due to the possibility of inaccurate and vague descriptions of analytical pipelines in scientific research and lack of access to the data and software. However, as a minimum standard, reproducing the research using the same data and analysis instructions helps to increase research reliability \cite{harris2018use} and enables testing other hypotheses without re-inventing the wheel. 

Reproducibility takes the form of {\em hard reproducibility}, which supports complete recreation of the data, software, and execution environment to arrive at the same result, or {\em soft reproducibility}, which allows the analytics task to execute on potentially different data (drawn from the same distribution), in different computational environments to arrive at statistically identical results. Hard reproducibility poses significant systems challenges by way of version control for data (data checkpoints), archiving data experiments, and recreating an identical computational environment (e.g., containerized software solution). As analytics kernels operate on increasing rate of heterogeneous datasets with complex dependencies across software components on ever changing hardware platform, hard reproducibility becomes an important though complex objective\cite{bui2017envisioning,goodman2016does}. In the context of computational science, scientific workflows are the artifacts that are being designed and published \cite{santana2015towards}. In terms of reproducibility, a workflow must be enriched by the description of the execution environment and the information of hardware and software components to be reproducible. In addition, to design a suitable hardware and software architecture, the scale, scope, and workload characteristics of big data analytics need to be considered, specifically for health data. For instance, by knowing the data access pattern of the users, it is possible to do memory hierarchy optimization to decrease latency and increase performance \cite{kambatla2014trends}.

We present a novel architecture for a software-hardware-data ecosystem using open-source tools and technologies such as Apache Hadoop \cite{hadoop} , JupyterHub\cite{jupyterhub} and Kubernetes\cite{kubernetes} with the goal of supporting (hard) reproducibility of analytic workflow with massive volume of heterogeneous data (e.g., SQL, time series, imaging). Apache Hadoop is one of the most popular technologies that serves massive amount of heterogeneous data in terms of storage and computation. Many organizations has adopted Apache Hadoop as their enterprise data warehouse to integrate heterogeneous data. For example, Geisinger\cite{davisgeisinger} as a healthcare provider has migrated to a big data enterprise data warehouse infrastructure based on Apache Hadoop to aggregate data from more than 100 sources.  JupyterHub is an open source multi-user data science user interface with support of multiple programming languages. Kubernetes is an open-source container orchestration platform to automate the management of the containerized applications across a distributed cluster for scalability and reproducibility. Although there are many systems that address reproducible workflow issues\cite{santana2015towards}, our focus is on reproducing the scientific benchmark code using Jupyter Notebook. However, our infrastructure has the potential to incorporate reproducible workflow system as a new service\cite{bisong2019kubeflow}. Our unique software-hardware-data ecosystem can be easily deployed in public and private cloud with minimal effort. It provides an analytic pipeline to aggregate heterogeneous data type (e.g., time series, images, structured data, among others) and is agnostic of programming language kernel (leveraging JupyterHub) and compliant with the HIPAA privacy law.

We evaluate the system using two large clinical datasets, MIMIC-III and Cerner Health Facts (CHF)\footnote{Data in Health Facts is extracted directly from the Electronic Medical Record(EMR) from hospitals in which Cerner has a data use agreement. Encounters may include pharmacy, clinical and microbiology laboratory, admission, and billing information from affiliated patient care locations. All admissions, medication orders and dispensing, laboratory orders and specimens are date and time stamped, providing a temporal relationship between treatment patterns and clinical information. Cerner Corporation has established Health Insurance Portability and Accountability Act-compliant operating policies to establish de-identification for Health Facts.}. Medical Information Mart for Intensive Care-III or MIMIC-III is a critical care data set curated with approximately sixty thousand intensive care unit admissions between 2001 to 2012 at Beth Israel Deaconess Medical Center\cite{johnson2016mimic,doi:10.1093/jamia/ocx084}. It includes both clinical and waveform data such as demographics, vital signs, laboratory tests and medications. CHF data is the clinical data associated with over 69 million patients for 19 years from 2000 to 2018. We evaluated the system with these data sets with regard to the performance and functionality with varying data volume and complexity of queries. 

The rest of the paper is organized as follows: the next section summarizes related works. Section \ref{architecture} provides an overview of the proposed architecture with a layered perspective. Section \ref{authentication} describes the solutions to provide secure access to the system to preserve privacy and security. Section \ref{data and evaluation} explains the data used to assess the performance and utilization of our system. Section \ref{discussion} describes the challenges in the current system that need to be addressed for large commercial enterprise solution. Finally, Section \ref{conclusion} concludes the paper.

\section{State Of The Art}
\label{related_work}
In this section we introduce several \emph{big data} platforms that have been developed to try to integrate complex health data and provide analytical solutions based on this data. We describe the architectures of these systems, their pitfalls as well as highlight the difference between these architectures and our system. 

\subsection{Patient-Centered Informatics Common: Standard Unification of Research Elements(PIC-SURE)} PIC-SURE\cite{bd2k,murphy2017grappling,GS2017Rcupcake} is an open source software platform to incorporate multiple heterogeneous patient level data including clinical, -omics and environmental data. The core idea of PIC-SURE is to utilize distributed data resources of various types and protocols such as SciDB, i2b2\cite{murphy2007architecture} and any other data systems by a single communication interface to perform queries and computations across different resources. For this purpose, PIC-SURE developed the \emph{Inter Resource Communication Tool (IRCT)}. IRCT is a resource-driven system and allows new resources to be integrated quickly. Furthermore, PIC-SURE API provides several pre-defined API resources that users can use to define and run a query and the results generated by a user can be available to that specific user only\cite{Jeremy2016BD2KPIC-SURE}. However, PIC-SURE API is not responsible for authentication and governance for individual access. While it provides a programming API that can be used in R and Python within an environment such as Jupyter Notebook, it is limited only to the pre-defined resources. In addition, to provide the reproducibility requirements (e.g. hard-reproducibility in terms of the same data and same environment) the proposed API should be integrated with a cloud-native development environment that users can develop and deploy their programs in a reproducible manner.   

\subsection{Informatics for Integrating Biology and the Bedside (i2b2)}
\begin{figure}[!t]
    \centering
    \includegraphics[width=2.5in]{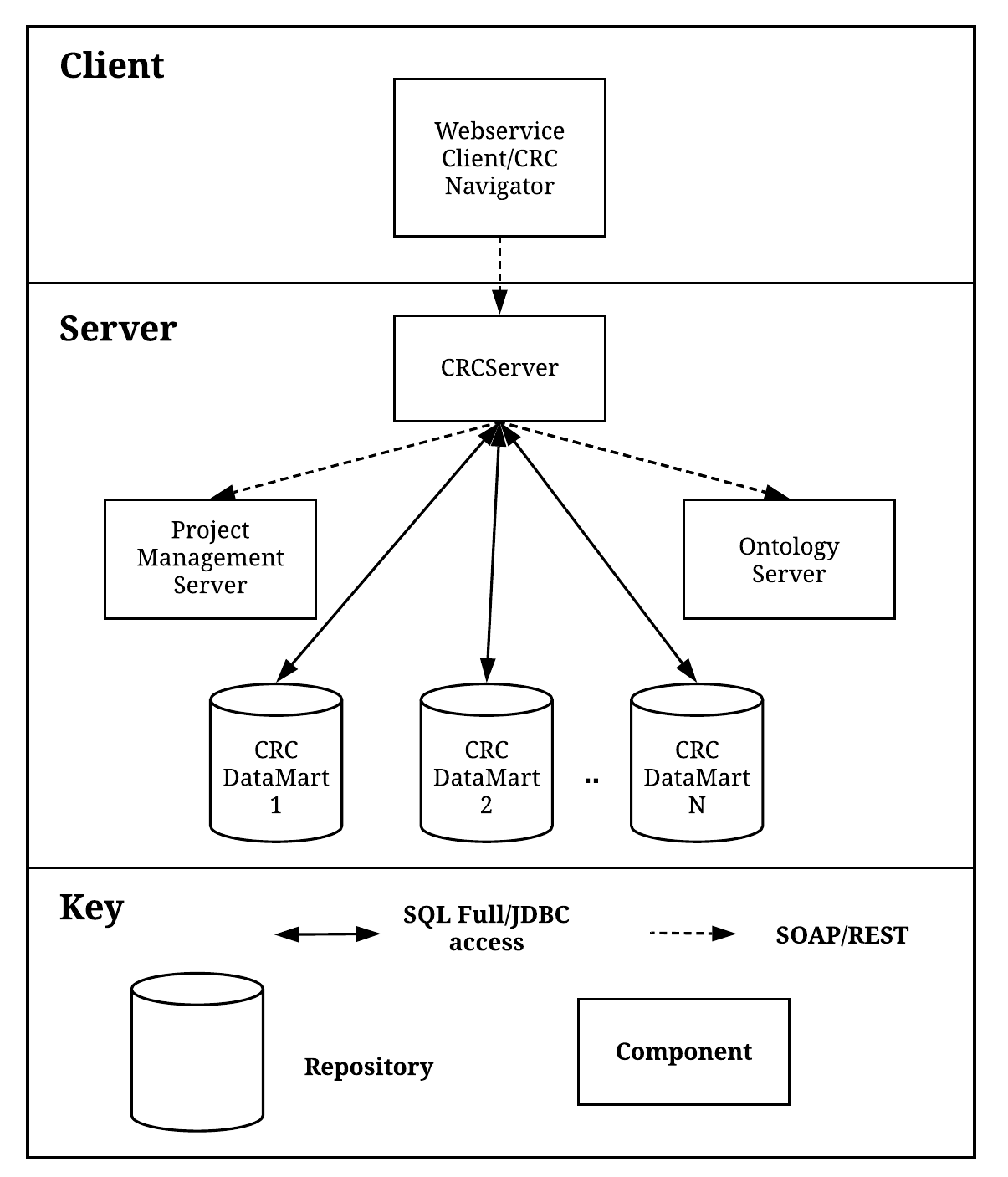}
    \caption{i2b2 Architecture\cite{murphy2007architecture}: The Component and Connector view shows a Client/Server view of i2b2's instances and the protocols they use for connection.}
    \label{fig:i2b2}
\end{figure}

i2b2\cite{murphy2007architecture} is an open source analytic query tool built on web services. i2b2 consists of a set of server-side software modules called \emph{Cell} and uses XML message for inter-cell communications which is illustrated in Figure \ref{fig:i2b2}.
Data is stored in a relational database such as Oracle using a common star schema data model. i2b2 is used by more than 200 healthcare institutions for cohort selection. Although using the relational database gives the advantage of SQL, with  web based architecture the proposed services are not as flexible as SQL itself. Furthermore, relational databases have the horizontal scalability issue and also are not optimized for large unstructured data. In the context of reproducibility, i2b2 allows users to share their queries within a group to be repeated with the same data or used for a new set of data. However, this notion of reproducibility is only at the query level and there is no mechanism to share statistical methods and analytics pipelines. Although i2b2 provides a set of pre-loaded machine learning and statistical algorithms, its web based architecture allows to develop more sophisticated and complex algorithms as new web services for advanced programmers or engineers.   

\subsection{Observational Health Data Science and Informatics (OHDSI)} 
\begin{figure}[ht]
   \centering
    \includegraphics[width=2.5in]{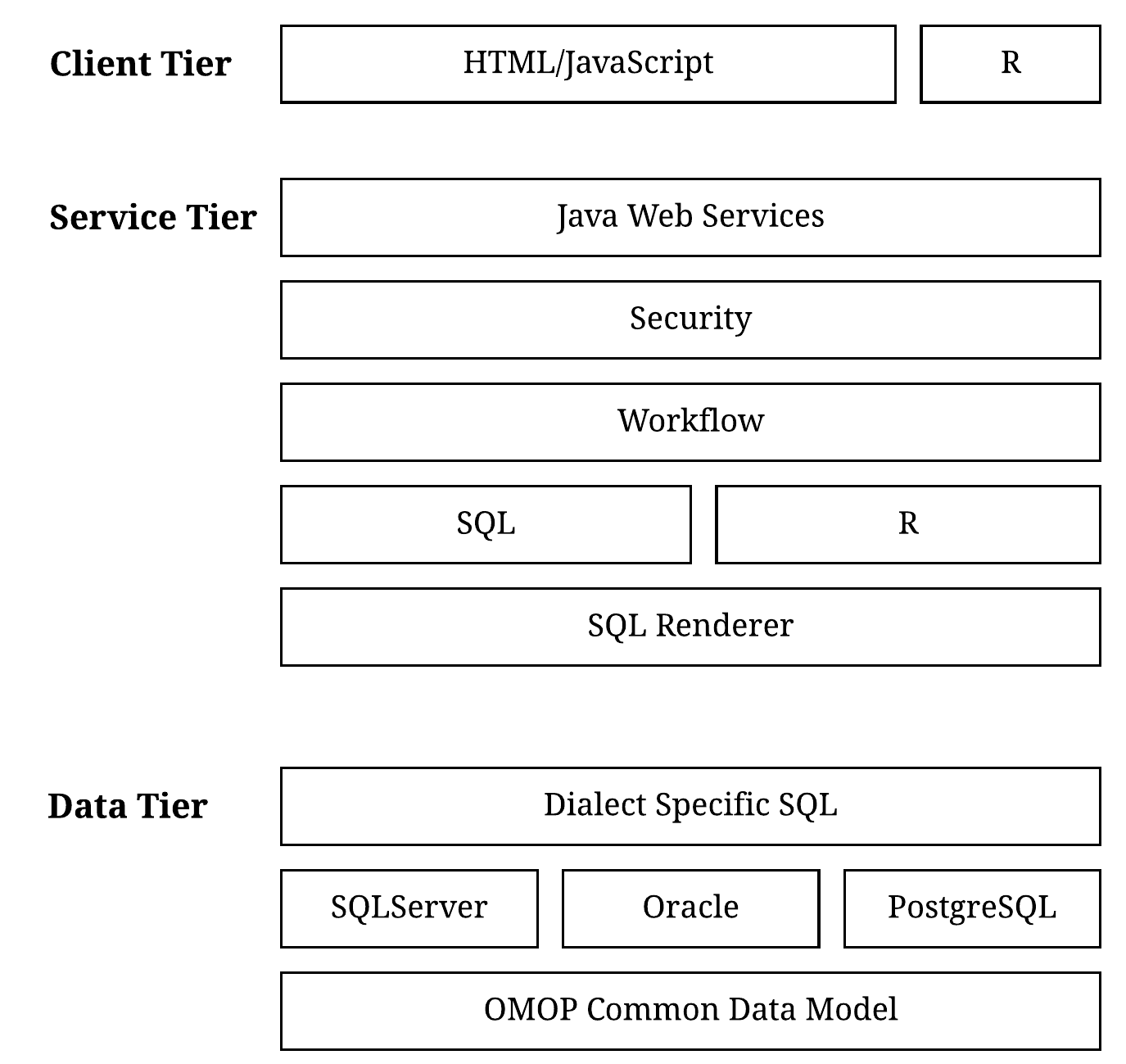}
    \caption{OHDSI Layered Architecture\cite{hripcsak2015observational}: This layered architecture shows a visual illustration of the main components of Client Tier, Service Tier and Data Tier of OHDSI's technology stack.}
    \label{fig:ohdsi}
\end{figure}

OHDSI\cite{hripcsak2015observational} is an open network of multiple observational data holders such as healthcare providers, hospitals, insurance companies, etc. It requires the network participants to translate their data into a common data model \emph{(OMOP\footnote{Observational Medical Outcomes Partnership})} in order to reuse the same query across different systems. Figure \ref{fig:ohdsi} shows the layered architecture for OHDSI consisting of three layers \emph{Client Tier}, \emph{Server Tier} and \emph{Data Tier}.
Similar to i2b2, OHDSI only works with relational databases such as Oracle and PostgreSQL and has the same challenges of being limited to structured data and issues with scalability. In terms of reproducibility, the OHDSI community has developed methods, libraries and tools consisting of R packages and shared them in the community's repository to be accessible by everyone in the community. 

\subsection{Unified Platform for Big Trajectory Data Management and Analytics (UlTraMan)}In \cite{DingXin:2018}, the authors propose a unified platform for big trajectory data management and analytics called UlTraMan. UlTraMan is an integrated platform of extended Apache Spark in both data storage and computing aspects. The extension has been made by integrating Spark with Chronicle Map and enhanced MapReduce. Chronicle Map is a key-value store for the purpose of data storage and random data access. They also have improved MapReduce by random data access optimization for computing purpose. The goal is to handle the pipeline of transforming, processing and analyzing the big trajectory data such as data generated by cars and mobiles.  
\begin{figure}[ht]
    \centering
    \includegraphics[width=2.5in]{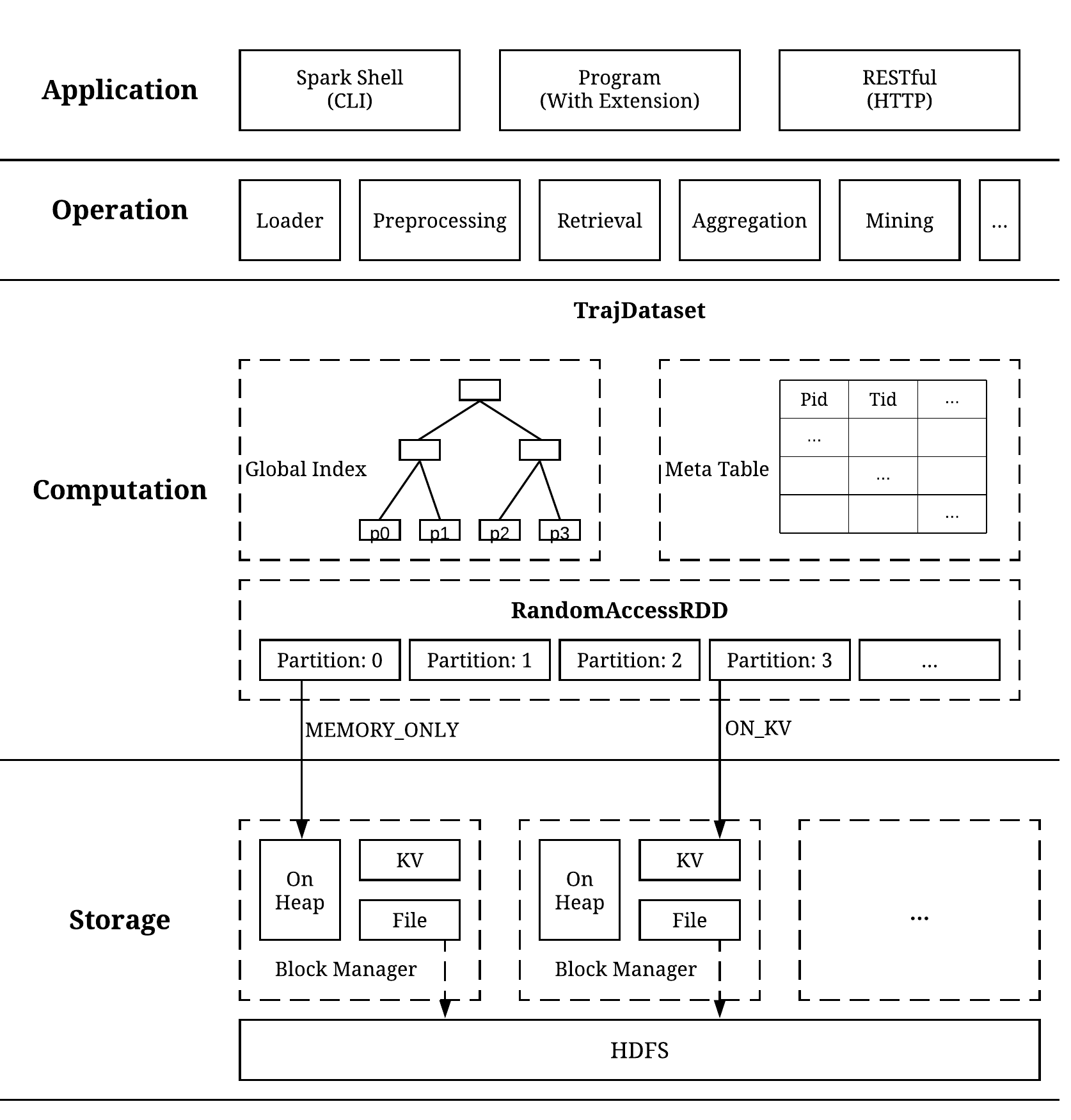}
    \caption{ULTraMan Layered Architecture\cite{DingXin:2018}: The Storage and Computation layers show the underlying unified engine of UltraMan which is an integration of Apache Spark, TrajDataset abstraction and Chronicle Map.}
    \label{fig:ultraman}
\end{figure}
Chronicle Map is a high performance, in-memory key-value data store that plays as internal block manager of Spark in this platform. To utilize random-access based techniques and optimization such as hash-map and indexes the authors improved MapReduce by an abstraction called TrajDataset. TrajDataset enables random access in both the local and global levels. The platform consists of four layers: storage, computation, operation, and application. The storage layer handles the data and the indexes. The computation layer is responsible for the distributed computations using the TrajDataset abstraction to enable random access. The operation layer supports a programming language interface to develop reusable components to analyze and process the data. In the application layer, UltraMan offers multiple ways of interaction for users, such as Spark shell and HTTP server for web requests. However, it doesn't propose any mechanism to provide reproducibility in terms of sharing and reproducing data pipelines and analytics. Also, UltraMan is not suitable for highly sensitive data such as health data. 

\subsection{WaveformECG}
Winslow et al. \cite{winslow2016waveformecg} developed WaveformECG, an open source web based platform that supports interactive analysis, data visualization, and annotation of the Electrocardiogram (ECG) data. ECG is a well known time series data type in cardiovascular research. It can contain high frequency data  and is primarily used for monitoring the heart condition or diagnosing diseases such as atrial fibrillation. 
\begin{figure}[ht]
    \centering
    \includegraphics[width=2.5in]{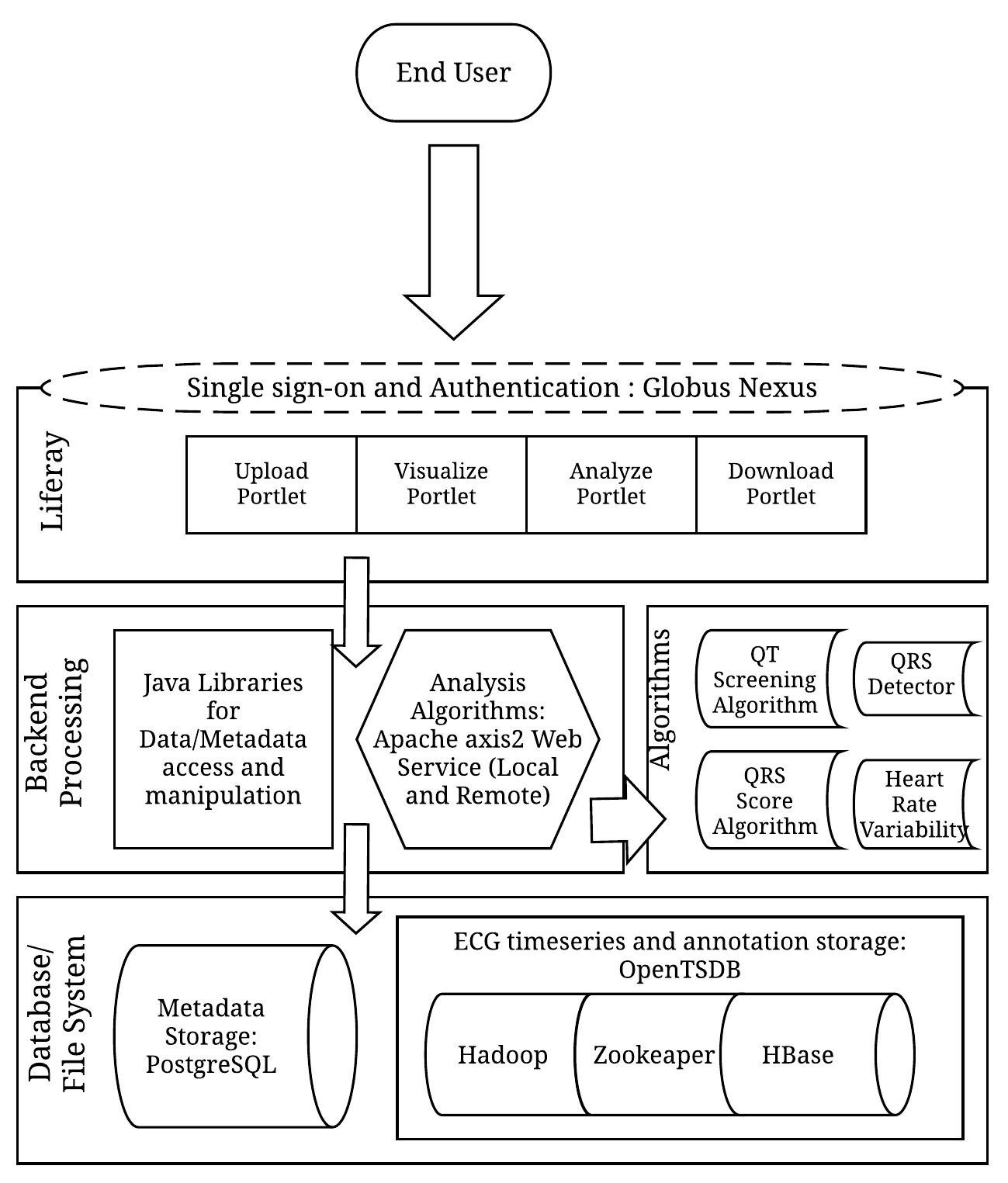}
    \caption{WaveformECG Architecture\cite{winslow2016waveformecg}: WaveformECG uses a web based architecture to provide access to data, analysis algorithms, analysis results and data annotation.}
    \label{fig:waveformecg}
\end{figure}
Users can login to WaveformECG through a portal developed using Liferay Portal Community Edition that is extended to use a federated identity provider Globus Nexus for authentication and authorization. After the authentication layer, there are four portlet interfaces to upload, visualize, analyze and download- supported by several backend libraries. The upload and visualization interfaces utilize OpenTSDB\cite{sigoure2012opentsdb}, which is an open source distributed time-series database with Apache Hadoop and Hbase. This architecture with Apache Zookeaper provides an interface to process real-time streaming ECG data. In addition, OpenTSDB provides RESTful APIs to access its storage and retrieve data that makes it possible to query ECG data from other software.  Analysis algorithms are available as web services accessed through Apache Axis2. When a user selects data file(s) and executes algorithm(s), data will be retrieved by a HTTP request from OpenTSDB and written in the desired format (e.g., XML or WFDB) by the algorithm. 
Visualization services let users examine the actual ECG data directly and annotate them manually. WaveformECG is integrated with i2b2 clinical data warehouse so the selected cohort in i2b2 can be sent to WaveformECG for further analysis.

\subsection{Starfish} The ability to perform cost-effective analysis in a timely fashion  over big heterogeneous data is one of the purposes of Hadoop software stack. Hadoop provides different parameters such as the number of map and reduce tasks that can be tuned based on the job specification for optimal performance. However, most of the Hadoop users lack the expertise needed to tune the system for good performance. Starfish \cite{chen2012interactive} is a self-tuning system for big data analytics that is built on top of Hadoop. It tunes the system based on the user's needs and the system workloads. Starfish tunes at three different  levels, \emph{Job}-level by approximating the job's statistics and performance mode,  \emph{Workflow}-level by handling the unbalanced data layout because of the data-local fashion in Hadoop and \emph{Workload}-level by optimizing workflows based on the shared data-flow or re-using the intermediate data and handing them to the workflow scheduler.

\subsection{Cerner's HealtheDataLab}
Cerner's HealtheDataLab\cite{ehwerhemuepha2020healthedatalab} is a cloud computing solution utilizing Fast Healthcare Interoperability Resources (FHIR) for data standardization, and distributed computing solutions for advanced data analysis. It is designed to serve the researchers to develop data analysis and machine learning models in a HIPPA compliant, high-performance and cloud-based computing environment. Jupyter Notebook is the front-end of the platform and it provides a web-based interface to develop, document, and execute codes in Python and R programming languages. Apache Spark is the core component of the backend as a computing engine. Apache Spark is an in-memory, parallel analytic engine that provides big data analysis and a rich machine learning libraries. HealtheDataLab is deployed in Amazon AWS to provide scalability and elasticity and data is stored in Amazon Simple Storage Solution (S3). The primary source of data is the  Cerner’s HealtheIntent platform.

\section{An Overview of the Proposed Architecture}
\label{architecture}
We provide an overview of the proposed architecture in Figure \ref{fig:Architecture}. The system consists of four layers: \textit{Infrastructure}, \textit{Storage}, \textit{Computation} and \textit{Service}. The Infrastructure layer represents the underlying hardware that we are using for the system. The Storage layer consists of the storage technologies. The Computation layer contains different distributed computation tools for different scenarios. Finally, the Service layer consists of different services in a container based environment. We explain each of these layers in detail in the following subsections.

\subsection{Infrastructure Layer}
\label{Infrastructure Layer}
The Infrastructure layer is the underlying equipment made up of a cluster of computers with high capacity in memory, storage and processors. This layer can be equipped with specific hardware such as specialized microprocessors including \emph{Graphical Processing Unit (GPU)} or \emph{Solid-State Drive (SSD)}. GPUs are capable of handling a few specific tasks in a very short time. SSDs are special disks that increase the performance for read/write (I/O) operations.

\subsection{Storage Layer}
\label{Storage Layer}
\begin{figure}[!t]
    \centering
    \includegraphics[width=\columnwidth]{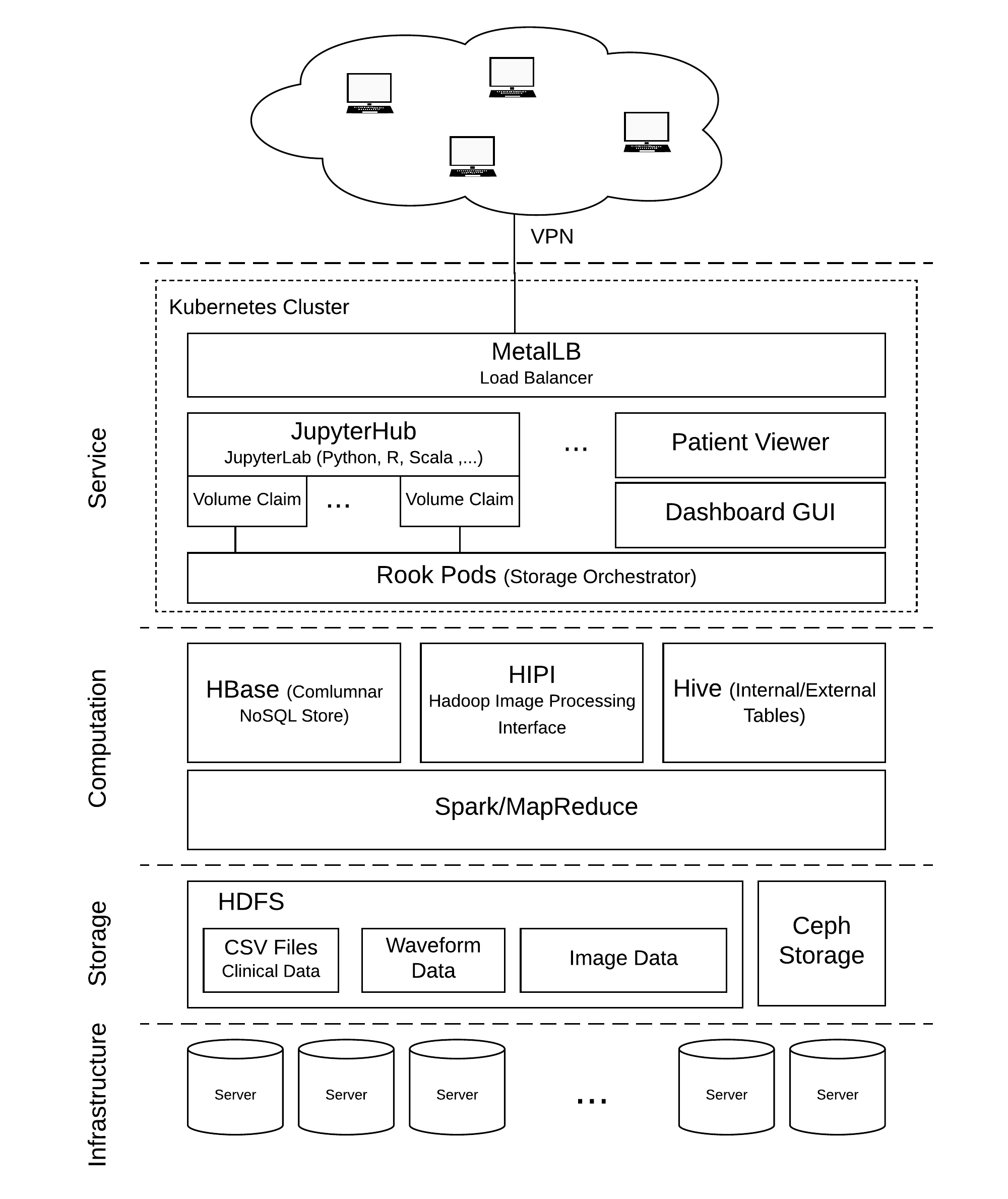}
    \caption{RCHE-HUB Layered Architecture: This layered architecture shows the integration of open-source cloud based technologies to provide a data-type agnostic, programming language agnostic, scalable and reproducible environment in a privacy-preserving manner suitable for highly sensitive data.}
    \label{fig:Architecture}
\end{figure}

The Storage layer is responsible for storing and managing large data with heterogeneous data types (e.g. images, time-series and structured data) of varying size. Figure \ref{fig:Architecture} shows this layer consists of two major components to keep the storage management separate for the data and users' files. The first component is \emph{Hadoop Distributed File System (HDFS)}. HDFS is the storage system used by Hadoop applications and provides high throughput data access for large and heterogeneous data sets. Different types of data such as structured, unstructured and image data can be stored on HDFS and be accessible by the components in computation layer. Different types of clinical data such as medical image data, time series data from medical devices such as ECGs, physiological monitoring devices, genomcis and clinical data can be stored and processed for analysis. 

The second component is Ceph \cite{weil2006ceph} storage which is directly used by the service layer. Services on a data processing environment are mostly stateful and hence need a place to store their intermediate data and files. For this purpose, we have used Ceph as the storage platform to provide persistent volume for services' persistent data. In order to ease the administrative overhead, Ceph storage is managed automatically by a Rook \cite{rook} cluster, which is deployed in the service layer. We explain the Rook cluster in more details in Section \ref{Service Layer}.

\subsection{Computation Layer}
\label{Computation Layer}
To utilize a cluster of computers and processors, Hadoop provides a framework that holds a collection of open-source software and tools. The computation layer holds the tools we  use to support the distributed processing and computing for a large amount of data. 

\emph{Hadoop Image Processing Interface (HIPI)} \cite{hipi} is an image processing library that can be used with Apache Hadoop MapReduce for parallel processing. HIPI helps to store a large collection of images on the HDFS and also makes them available for distributed processing. Furthermore, we can use HIPI  with well known open source image processing libraries such as OpenCV that provides a variety of computer vision algorithms.

To store the temporal and waveform data, we use HBase (non-relational database) that runs on top of HDFS. HBase stores data as key/value pairs in columnar fashion and provides real-time read and write access with low latency. However, executing queries against HBase are not convenient for a relational data schema. To support SQL-like queries for relational database schema (e.g., for electronic health records (EHR)) we use Apache Hive. Apache Hive is a data warehouse software on top of Hadoop and provides SQL-Like queries (i.e. HiveQL). Hive, by default uses Hadoop MapReduce to execute queries and analyze the data. MapReduce performs the processing in disk which is I/O intensive and very slow. Hive can also use other distributed computation engines such as Apache Spark. In order to improve the performance and processing speed, we use Apache Spark on top of Hive. The biggest advantage of Spark over MapReduce is that it performs the processing in-memory and in parallel using \emph{Resilient Distributed Dataset (RDD)}.  This improves  performance due to low disk communication needs.

\subsection{Service Layer}
\label{Service Layer}
The Service layer provides a container based environment for scalability, self-healing, auto-managing and monitoring services running in containers (\emph{micro-services}). A popular open-source container technology is Docker\cite{docker}, which allows us to create, run and manage containers on a single operating system. However, for the case of a cluster of hosts, it is hard to keep track of all of the containers on different hosts. In such a  scenario, we can leverage the open-source container orchestration platform Kubernetes\cite{kubernetes} to automate the management of the containerized applications across the cluster of hosts.

Kubernetes was originally developed to support stateless applications with no need for storage. In order to deploy applications that need to store their data in persistent storage (known as \emph{stateful} applications) on a Kubernetes cluster, we need storage that is available anytime and anywhere the containers can be deployed. Cloud providers offer their own storage services to provide persistent volume for persistent data. Unfortunately, for on-premise systems with Kubernetes, we cannot rely on these convenient storage services. In order to address this issue we use the storage orchestrator Rook. Rook \cite{rook} is  an open-source  cloud-native  (container based)  storage  orchestrator that takes advantage of the underlying container based environment for Kubernetes to facilitate managing, monitoring, scaling, provisioning, healing, and recovery of the storage services. Rook supports multiple storage solutions such as Ceph, \emph{Network File System (NFS)} and Cassandra to be integrated with cloud-native environments.  For a production environment the Ceph storage system is recommended since it is more stable since most other solutions are still in Alpha or Beta versions. Ceph is a highly scalable distributed storage system that provides block storage, object storage, and shared file systems. We use block storage that can be mounted to a single pod to store its persistent data and shared file system which is shared between multiple pods.

Another issue to address for on-premise installation of Kubernetes is routing traffic into the cluster using load balancers. Public clouds such as GCP or AWS have convenient services for routing traffics to Kubernetes cluster. However, most of the standard load balancers are only deployable on public cloud providers and are not supported for on-premise installation. Fortunately, MetalLB\cite{metallb} has been developed to address this issue. MetalLB is an on-premise load balancer that provides two different configuration, BGP based and Layer2 based load balancing.  MetalLB is mostly responsible for distributing the network or application traffic across multiple servers to increase the capacity and reliability of the applications.

Such a scalable container based environment that handles storage and service load balancing is ready to deploy services and applications in a reproducible manner. One of the primary services we provide for researchers in our platform is JupyterHub\cite{jupyterhub}, an environment for developing applications to analyze and process data. JupyterHub is an open source multi-user web based programming language interface that supports multiple programming language Kernels such as Python, R, Scala. Deploying JupyterHub on the Kubernetes cluster makes it manageable and scalable. Using JupyterHub helps to use single server JupyterLab for a group of people. Every instance of a JupyterLab will be deployed inside a docker container in our Kubernetes cluster with two different spaces as storage. A block storage is used as a local directory and a shared file system which is shared between all JupyterLab instances. In addition, utilizing the containerization helps to provide reproducible applications which can be shared easily among the researchers.

\section{Authentication}
\label{authentication}
Collecting personally identifiable health information is important to be able to conduct research. However, protecting  personal privacy is essential. For example, the HIPAA privacy law mandates that certain individual information cannot be disclosed by researchers \cite{ness2007influence}. To that end, the privacy and security of patients health information is a top priority of every health-related organization \cite{winn2001confidentiality}. To guarantee the safety of sensitive data, we provide multiple security layers in our cluster. First, all the data are de-identified and all personal information are removed according to our contracting with the data providers. In order to access the data, a user needs to sign a confidentiality agreement. Furthermore, all the connections to the cluster are through a secure VPN with two factor authentication. In addition, JupyterHub has support for user authentication via a third-party OAuth provider, including Google, GitHub, and CILogon\cite{cilogon}. We use CILogon that supports Purdue University as an identity provider. This helps us to take advantage of the Purdue's 2-factor authentication. 
In addition to the aforementioned techniques, we utilize Apache Ranger\cite{ranger} to authorize different levels of access. Apache Ranger is a framework to manage and monitor data security across the Hadoop cluster, and allows us to define different types of user-level allow/deny permissions for each data table and database. 

\section{Data and Performance Evaluation}
\label{data and evaluation}
\subsection{Experimental Setup}
\begin{figure}[t]
    \centering
    \includegraphics[width=\columnwidth]{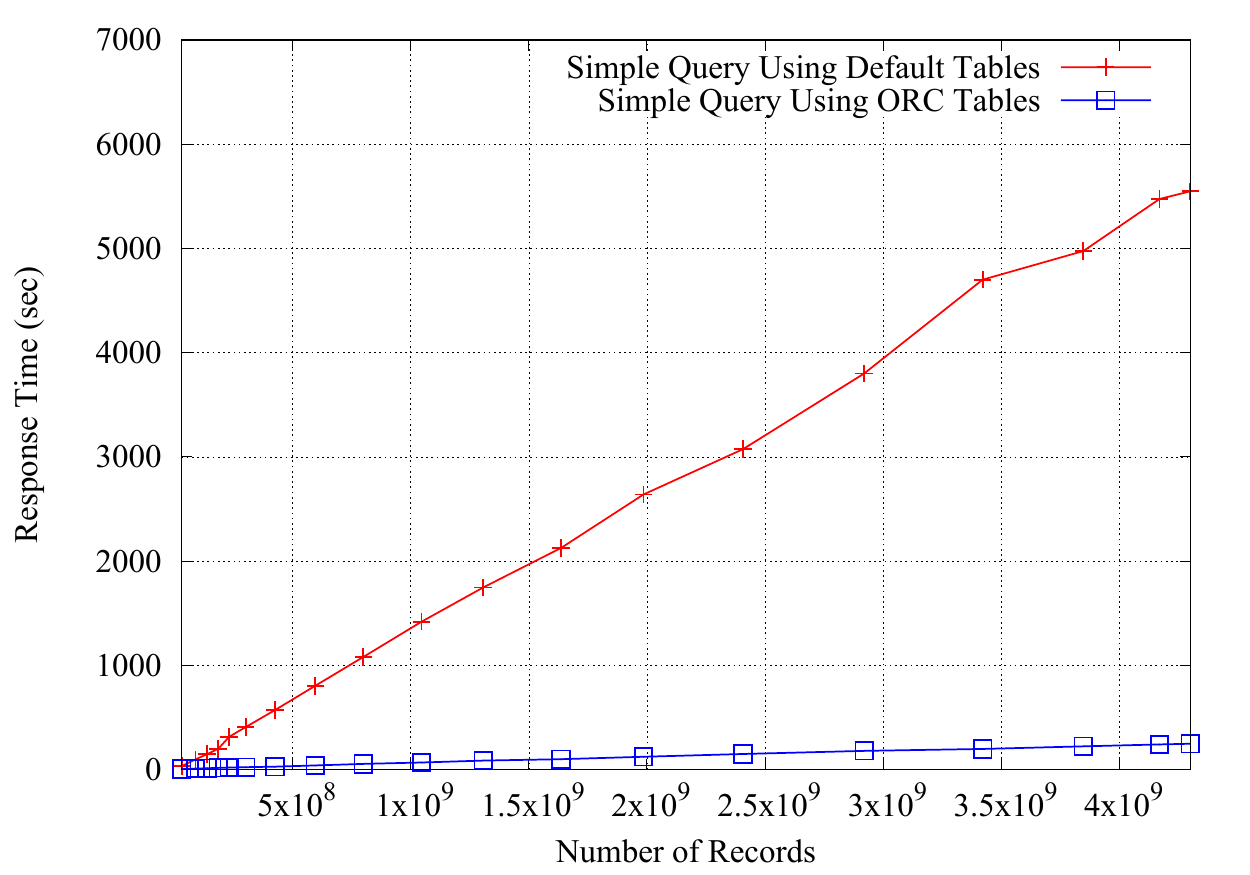}
    \caption{Response time vs. number of records for a simple query using default and ORC tables: The response time increases as we increase the size of the data for both types of tables. ORC tables show a great improvement in response time using the optimize way of data storing.}
    \label{fig:rtime-simple}
\end{figure}
We have deployed our architecture with an on-premise private cloud system at the Regenstrief Center for Healthcare Engineering (RCHE) at Purdue University. The system has 10 nodes in total, 8 worker nodes and 2 master nodes, to perform the performance evaluation. Each worker node is equipped with $188$ GB memory, $24$ processors each with $12$ cores and $30$ TB disk space. Each master node is equipped with $250$ GB memory and $40$ processors with $10$ cores for each processor. We use HDP-2.6.5.0 for the Hadoop cluster with Apache Spark version $2$, and client version 1.15.0 for the Kubernetes cluster.  

\subsection{Datasets}
Cerner Health Facts (CHF) data is a clinical database that includes diagnostic information, demographics, medical history, admissions, discharges, drug prescriptions, and laboratory tests associated with over 69 million patients for the 19 year period of 2000 to 2018. This longitudinal data for individual patients comes from the electronic health records (EHR) of over $750$ hospitals . CHF is HIPAA compliant and de-identified and is used by selected community and academic facilities across the United States.  Purdue University has Data Use Agreements to use the data for research purposes. The use of the CHF data has been approved by the Purdue University Human Research Protections Program (HRPP) Institutional Review Board (IRB) with an exemption determination (PROPEL Determination No. 29007411). The dataset has previously been used in other environments to answer some specific clinical questions \cite{petrick2016temporal}. However, no such computational platform as described here have been developed and implemented for the CHF dataset\cite{miao2018assessment}.   \\
A second source of data,  MIMIC III and associated Physionet tools, is a publicly available system for EHR data and high resolution time series data from medical devices. It is created and maintained by the Laboratory for Computational Physiology at MIT \cite{johnson2016mimic,doi:10.1093/jamia/ocx084}. The database contains high resolution waveform data and clinical information on patients admitted to the Intensive Care Unit (ICU) since 2001 at the Beth Israel Deaconess Medical Center. 

\begin{figure}[t]
    \centering
    \includegraphics[width=\columnwidth]{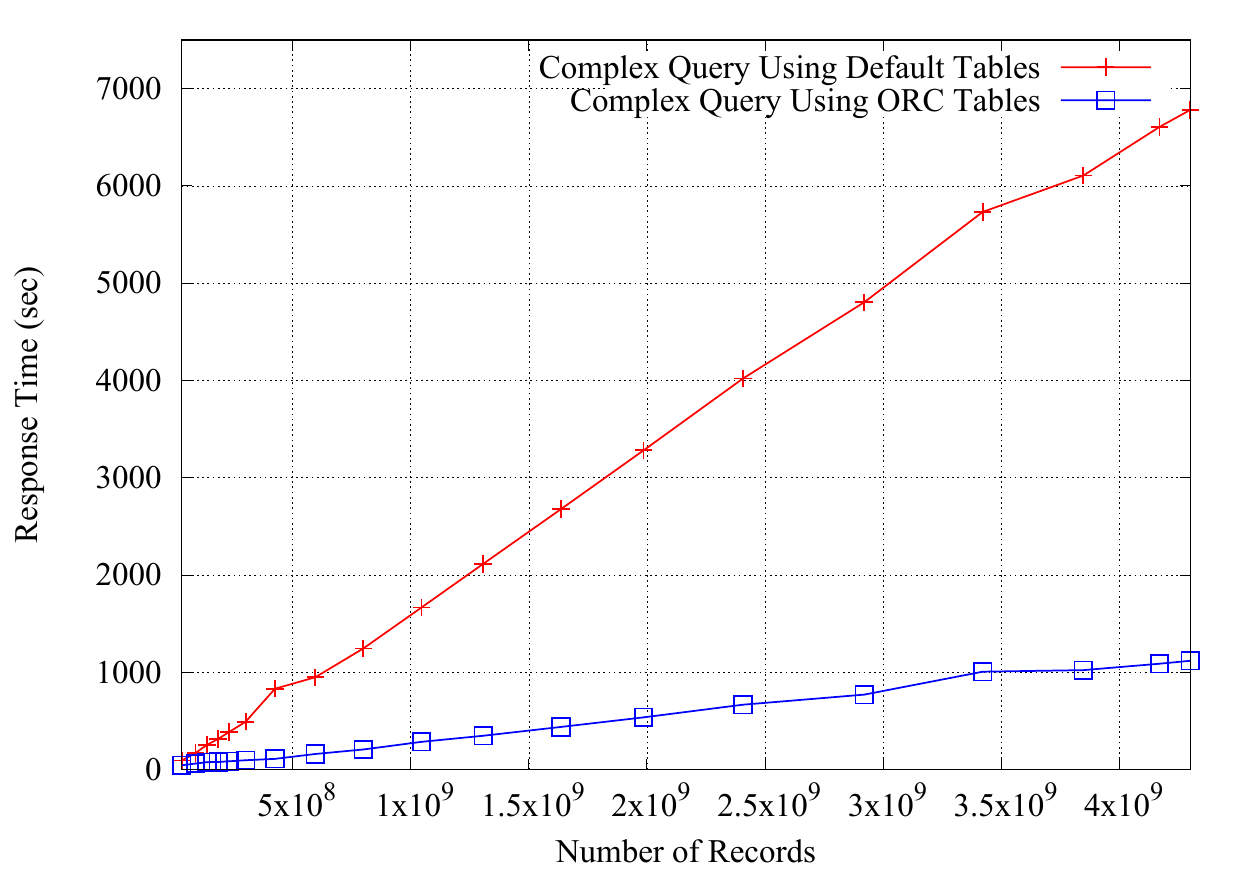}
    \caption{Response time vs. number of records for a complex queries using default and ORC tables: The response time increases as we increase the size of the data for both types of tables. Even for complex queries containing multiple Join operations and aggregation functions, ORC tables show a great improvement in response time using the optimized way of data storing. }
    \label{fig:rtime-complex}
\end{figure}

\subsection{Performance Evaluation}
\begin{figure}[t]
    \centering
    \includegraphics[width=\columnwidth]{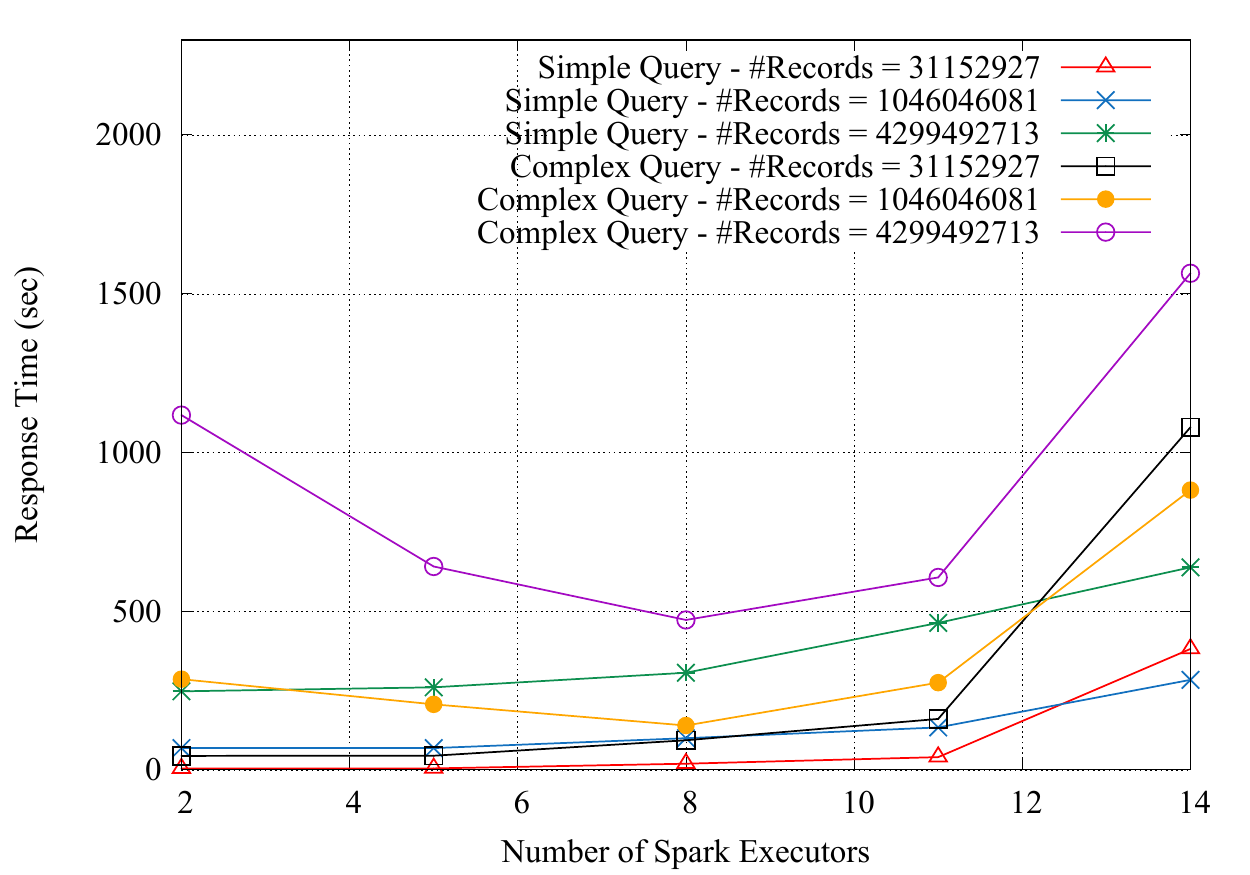}
    \caption{Response time vs. number of Spark executors using ORC tables for simple and complex queries with various sizes of data , spark-executor-mem = 256M - spark-executor-cores = 3: There is a trade-off for increasing the number of Spark executors, namely, as it increases the parallelism also increases the overhead from communications and garbage collection.}
    \label{fig:rtime-exec1}
\end{figure}

We evaluate the functionality and performance of the proposed architecture from different perspectives. As the performance metric, we measure the response time for a query which is submitted from the JupyterLab instance on the  Kubernetes cluster to the Hadoop cluster. To capture the response time, we used function \emph{time()} from the "time" library in Python, which returns the number of seconds that has passed since epoch (for Unix system epoch has started from January 1, 1970, 00:00:00). By measuring the time before and after submitting the job to Hadoop and calculating their difference, we obtain the approximate response-time in seconds.  For the benchmark, we chose tables \verb|Encounter| and \verb|Lab_procedure| from CHF. The \verb|Lab_procedure| table is one of the largest tables in CHF and has information on lab events, and the \verb|Encounter| table has information on events associated with each patient and is linked to \verb|Lab_procedure|. The tables are stored in HDFS as CSV files but are accessible through Hive external tables. In this case, the default format of Hive tables are TEXTFILE, but we also created Optimized Row Columnar (ORCFILE) tables which stores the collection of rows in one file in a columnar way.  Therefore, specific columns can be accessed faster and in parallel. Furthermore, we define both a simple and a complex query. The simple query is an aggregation function to simply count the number of records in the \verb|Lab_procedure| table. The complex query joins two tables, categorizes a specific lab result value, and gives the distribution of number of patients over the categories. Using the different types of tables and queries, we define the following three scenarios:
\begin{enumerate}
    \item \textbf{Scenario One: Simple Query} In this scenario we measured the response time for the simple query to count the number of records against both types of Hive tables for various sizes of data. As the results in Figure \ref{fig:rtime-simple} show, response time goes up as the size of the data increases but the increase in latency is significantly slower using ORC tables. The ORC file groups the data rows in stripes along with indexes, it improves the performance when Hive processes the data. 
    \item \textbf{Scenario Two: Complex Query} In this scenario, we considered the complex query that joins two tables, categorizes a specific lab result value, and gives the distribution of number of patients over the categories to analyze the cluster using both types of tables for different sizes of data. The trend is similar to  scenario one, but in overall we have a larger response time because of the complexity of the query and the joining of two large tables (Figure \ref{fig:rtime-complex}).
    \item \textbf{Scenario Three: Spark Parameter Optimization} We evaluate the impact of the number of Spark executors on the response time. Since the ORC tables has shown a drastic improvement on the response time, we decided to use it as the base table format to evaluate the impact of the Spark parameters on the performance. As Figure \ref{fig:rtime-exec1} shows, increasing the number of Spark instances  doesn't improve the response time for the small data size independent of the complexity of the query.  Increasing the number of executors for large data and complex queries (i.e. Complex Query - \#Records = 1046046081 or Complex Query - \#Records = 4299492713) improves the response time. However, after a specific number of executors, the response time increases due to the overhead of distributed communication.   
\end{enumerate}
In order to tune the Spark parameters to achieve optimal performance, a user should be aware of the size of the data, distribution of the data across the cluster and the complexity of the query. Otherwise, simply increasing the number of executors, amount of memory or the number of virtual cores for each executor would not necessarily lead to an improvement. For example, having executors with large amounts of memory often leads to excessive garbage collection overhead. Further, running executors with a single core and small memory suitable for only one task doesn't allow for the benefit of parallelism.

\section{Discussion}
\label{discussion}
We provide a novel multi-layer cloud based infrastructure using Hadoop and Kubernetes clusters for the purpose of addressing issues of scalability, heterogeneity, and reproducibility. This architecture allows users to easily implement, deploy and share their research applications in different programming languages. For the system administrator, the proposed architecture is equipped with managing and monitoring tools that facilitates scaling, managing, and monitoring both the data and the applications for security and authentication. This architecture allows us to develop and test performance issues related to data aggregation, query processing time, container deployment, and management, with health data. Using Kubernetes as a Docker orchestration platform helps to develop new functionalities as light and salable micro-services with no dependency to the underlying system. 

In the future, we will address some critical aspects of the system to better support collaborative research infrastructure for health data. For example, while preserving privacy is one of the critical challenges in working with health data, de-identifying the data might cause information loss for analytic solution resulting in less accurate methods. Consequently, it is crucial to provide different levels of access so that researchers can perform robust and accurate analysis while at the same time ensure the data privacy of individuals. It is also important to have an audit trail for each user that can be processed and analyzed with advanced artificial intelligence methods for any unauthorized activities. 
In addition, such a platform is usually used by physicians, computational scientists and health-related researchers who typically have limited training in computational systems and approaches. Therefore, providing a minimum set of simple abstractions for different stages of a machine learning workflow from data pre-processing, loading, and model training will be beneficial to the research community.

In addition, JupyterLab does not currently provide a collaborative programming work-space in the cluster environment, rather it provides an isolated environment for each researcher to work on their projects in a single server JupyterLabs. Here, developing an extension to JupyterLab that can leverage the shared file system to define different levels of access to the shared projects would be of value. However, this will add challenges for managing the cluster in a shared work space. 

Exception handling for driver failure in the client side is another important feature to develop. In a client mode Spark job, the driver component is created on the client side, and it is responsible of creating, and planning the job to be executed by the executors in parallel. In this mode, if the driver encounters any exception, the computation will fail and the user will have to restart the computation. Since Apache Hadoop provides fault tolerance in terms of resuming computations in case of software, hardware and node failure, we will implement exception handling due to driver failure resuming user's job in case of crash on the client side. 

Finally, in terms of data visualization in a big-data environment with petabytes of data, we will implement distributed data visualization functionalities using Apache Spark by providing an abstraction of the data chunks handled by each executor to the driver.

As another future plan, we will also evaluate the performance of the architecture with other types of data such as time series and image data. Considering the high potential of our system, we plan to make an improvement by designing and developing new cloud-native sub-systems to increase the performance in a practical way. To evaluate the future optimization techniques, we will compare the performance of our system with the baseline SQL databases in case of relational datasets such as clinical data.

 \section{Conclusion}
\label{conclusion}
In this paper, we proposed a novel architecture for a healthcare software-hardware-data ecosystem using open source tools and technologies including Apache Hadoop, Kubernetes and JupyterHub. Our base architecture can store, manage and process large heterogeneous data in a reproducible manner and it has potential to be extended by new sub-systems in a cloud-native way. We also evaluated the system with different scenarios and with large clinical data sets. For baseline performance evaluation, we compared two types of data management using Apache Hive (Text and ORC) to illustrate the difference. We also plan to evaluate the system using other types of data such as time-series and image data.

\section*{Acknowledgments}
We would like to thank the Information Technology at Purdue (ITaP) department for their support in managing the security, networking and operating system. Specifically, we appreciate the genuine support from Pascal Meunier and Andrew Thomas.

\bibliographystyle{ACM-Reference-Format}
\bibliography{ref}

\appendix

\end{document}